\documentclass[prd,superscriptaddress,altaffilletter,showpacs]{revtex4}
\usepackage{amssymb}
\usepackage{amsmath}
\newcommand{\be}{\begin{equation}}
\newcommand{\ee}{\end{equation}}

\newcommand{\ba}{\begin{eqnarray}}
\newcommand{\ea}{\end{eqnarray}}

\begin{document}

\title{Localization of gauge fields in a tachyonic de Sitter thick braneworld}

\author{Alfredo Herrera--Aguilar}\email{aherreraaguilar@mctp.mx}
\affiliation{Mesoamerican Centre for Theoretical Physics, Universidad Aut\'onoma de Chiapas,\\
Ciudad Universitaria, Kil\'ometro 4, Carretera Emiliano Zapata, Colonia Real del Bosque (Ter\'an), CP 29050, Tuxtla Guti\'errez, Chiapas, M\'{e}xico.}
\affiliation{Departamento de F\'{\i}sica, Universidad Aut\'onoma Metropolitana Iztapalapa,
San Rafael Atlixco 186, CP 09340, M\'exico D. F., M\'exico.}
\affiliation{Instituto de F\'{\i}sica y Matem\'{a}ticas, Universidad Michoacana de San Nicol\'as
de Hidalgo,\\Edificio C--3, Ciudad Universitaria, CP 58040, Morelia, Michoac\'{a}n, M\'{e}xico.}

\author{Alma D. Rojas}\email{arojas@mctp.mx}
\affiliation{Mesoamerican Centre for Theoretical Physics, Universidad Aut\'onoma de Chiapas,\\
Ciudad Universitaria, Kil\'ometro 4, Carretera Emiliano Zapata, Colonia Real del Bosque (Ter\'an), CP 29050, Tuxtla Guti\'errez, Chiapas, M\'{e}xico.}
\affiliation{AHEP Group, Instituto de F\'{\i}sica Corpuscular Ð C.S.I.C./Universitat de Val\`encia,\\ 
C/ Catedr\'atico Jos\'e\ Beltr\'an, 2, E--46980 Paterna (Val\'encia), Spain.}

\author{El\'\i \ Santos}\email{esantos@mctp.mx}
\affiliation{Mesoamerican Centre for Theoretical Physics, Universidad Aut\'onoma de Chiapas,\\
Ciudad Universitaria, Kil\'ometro 4, Carretera Emiliano Zapata, Colonia Real del Bosque (Ter\'an), CP 29050, Tuxtla Guti\'errez, Chiapas, M\'{e}xico.}

\begin{abstract}
In this work we show that universal gauge vector fields can be localized on the recently proposed 5D thick tachyonic braneworld which involves a de Sitter cosmological background induced on the 3--brane. Namely, by performing a suitable decomposition of the vector field, the resulting 4D effective action corresponds to a massive gauge field, while the profile along the extra dimension obeys a Schr\"{o}dinger-like equation with a P\"{o}schl-Teller potential. It turns out that the massless zero mode of the gauge field is bound to the expanding 3--brane and allows us to recover the standard 4D electromagnetic phenomena of our world. Moreover, this zero mode is separated from the continuum of Kaluza-Klein (KK) modes  by a mass gap determined by the scale of the expansion parameter. We also were able to analytically solve the corresponding Schr\"{o}dinger-like equation for arbitrary mass, showing that KK massive modes asymptotically behave like plane waves as expected. 

\end{abstract}



\maketitle

\section{Introduction and review of the thick tachyonic braneworld}

The original proposal that extra dimensions can be non--compact \cite{RubakovPLB1983136,VisserPLB1985,Randjbar-DaemiPLB1986} or even large \cite{AntoniadisPLB1990,ADD}  supplied new insights when attempting to solve the long standing cosmological constant problem \cite{RubakovPLB1983136,Randjbar-DaemiPLB1986,CosmConst} and the intriguing gauge hierarchy problem \cite{ADD,gog}.
Subsequently, after the surprising success of the Randall--Sundrum models in addressing the mass hierarchy and 4D gravity localization problems \cite{rs}, during last years, several braneworld configurations have been proposed in order to address different open problems in modern physics. 
Most of the braneworlds that generalize the original Randall--Sundrum setup make use of bulk scalar fields. In this framework, different kinds of scalar fields give rise to a wide range of scenarios 
\cite{De_Wolfe_PRD_2000,KehagiasTamvakis,Csaki_NPB_2000,dSbranes,fRwaveBW,ariasybarbosaetal,SR,Bazeiaetal,koleykar}. In particular, some authors have choosen a bulk tachyonic scalar field in order to address problems like the gauge hierarchy, the stabilization of the corresponding brane separation, and the localization of gravity and other matter fields \cite{Bazeiaetal, koleykar, senguptaetal1, germanetal}. However, many attempts to solve the highly non--linear field equations that arise in these scalar--tensor system lead to imaginary tachyon field configurations \cite{palkar,carrilloetal} or to delocalization of gravity and other kinds of matter, like scalar and gauge vector fields (see, for instance, the tachyonic braneworlds constructed in \cite{koleykar,palkar} and references therein).

An interesting issue within the braneworld paradigm is related to the question of whether bulk matter fields can be
confined to the brane by a natural mechanism. Usually, scalar fields and gravitons, as well as fermions, can be localized on branes of different types. 
However, within most of the so far proposed thin braneworlds, gauge vector fields cannot be localized on the 3--brane that represents our Universe by a gravitational trapping mechanism \cite{DvaliShifman,BajcGabadadze,Pomarol,Oda}. 
Quite recently some new braneworlds endowed with different mechanisms for localizing gauge vector fields, either on Weyl thick branes \cite{Liu0708}, on purely geometrical expanding (de Sitter) branes \cite{Guoetal2013} or in anisotropically curved branes supported by  phantom \cite{GogMid1} and real scalar fields \cite{GogMid2}, were constructed.
In particular, vector fields can be localized on the RS model in some higher--dimensional cases \cite{OdaPLB2000113}. 

In this paper we use a recently proposed thick tachyonic braneworld configuration with a de Sitter metric induced on the brane and a decaying warp factor that enables the localization of 4D gravity \cite{germanetal}, and show that universal gauge vector fields can also be localized in it. Moreover, when analyzing the dynamics of the vector field fluctuations, we see that their spectrum contains a {\it single} bound state corresponding to the 4D massless photon of the model, separated from a continuum of massive KK excitations by a mass gap. 

On the other hand, this kind of de Sitter 3-branes embedded in a higher dimensional manifold are also interesting from the point of view of cosmology. Since the aforementioned field configuration describes a 3--brane with 4D de Sitter symmetry (i.e. an expanding $dS_4$ brane) embedded in a positive definite and asymptotically Minkowski spacetime $M_5$, it qualitatively describes some aspects of the inflationary period of our own Universe \cite{guth,linde}. Moreover, since the second derivative of the scale factor with respect to time is positive, it can describe some aspects of the accelerated expansion of the Universe, related to its present epoch. It is known that the Cosmic Inflation theory is in accordance with the temperature fluctuation properties observed in the Cosmic Microwave Background Radiation. This agreement and other observational successes, together with the fact that inflation likely occurs at very high temperatures and its study involves relevant assumptions regarding the physical phenomena that take place at such high energies, lead to several serious attempts to find inflationary configurations within the framework of string theory and supergravity \cite{popeetal,burgess,SC}. 

Coming back to the tachyonic braneworld model, let us start by considering the following 5D action:
\begin{equation}
S = \int d^5 x \sqrt{-g} \left(\frac{1}{2\kappa_5^2} R - \Lambda_5\right) - \int d^{5}x \sqrt{-g}
V(T)\sqrt{1+g^{MN}\partial_{M} T\partial_{N} T}. 
\label{accion}
\end{equation}
This action describes gravity with a bulk cosmological constant $\Lambda_5$ minimally coupled to the bulk tachyon field $T$ \cite{bergshoeff,sen}, where $\kappa_5$ is the 5D gravitational coupling constant, $M,N=0,1,2,3,5$; and $V(T)$ denotes the tachyon self--interaction potential.

The second therm in (\ref{accion}) was originally proposed in \cite{bergshoeff} when studying a tachyon field living on the world volume of a non--BPS brane and has found interesting applications in braneworld cosmology \cite{mazumdar} and in string cosmology \cite{gibbons}. In \cite{padmanabhan} it was realized that this action can, in fact, be used for any relativistic scalar field. Moreover, solar system constraints were analyzed in \cite{devi} by using this tachyonic action as a scalar tensor theory.

We shall consider a warped 5D metric with a spatially flat cosmological background induced on the 3--brane:
\begin{eqnarray}
 ds^2 &=& g_{MN}dx^{M}dx^{N}
         = e^{2f(w)} \left[\hat{g}_{\mu\nu}(x)dx^\mu dx^\nu + dw^2 \right]
                \nonumber \\
       &=& e^{2f(w)} \left[- d t^2 + a^2(t) \left( dx^2 + dy^2 + dz^{2}\right) + dw^2\right],
       \label{metric_w}
    \end{eqnarray}
where $f(w)$ is the warp factor, $a(t)$ is the scale factor of the 3--brane, $\hat{g}_{\mu\nu}(x)$ denotes the 4D metric tensor ($\mu,\nu=0,1,2,3$) and $w$ is the fifth coordinate.

Here we shall not elaborate on how the solution of this system was obtained, we will simply present it and refer the reader to \cite{germanetal} for details. Thus, the solution for the metric ansatz, i.e. for the scale and warp factors, respectively reads 
\begin{equation}
a(t)=e^{H\,t}, \qquad  \qquad  f(w)=-\frac{1}{2}\ln\left[\frac{\cosh\left[\,H\,(2w+c)\right]}{s}\right],
\label{scalewarpfactors}
\end{equation}
the tachyon scalar field has the form
\begin{eqnarray}
T(w) &=& \pm\sqrt{\frac{-3}{2\,\kappa_5^2\,\Lambda_5}}\
\mbox{arctanh}\left[\frac{\sinh\left[\frac{H\,\left(2w+c\right)}{2}\right]}
{\sqrt{\cosh\left[\,H\,(2w+c)\right]}}\right]  \nonumber \\
&=& \pm\sqrt{\frac{-3}{2\,\kappa_5^2\,\Lambda_5}}\ \ln\left[
\mbox{tanh}\left(\frac{H\,\left(2w+c\right)}{2}\right)+\sqrt{1+\mbox{tanh}^2\left(\frac{H\,\left(2w+c\right)}{2}\right)}
\right] , 
\label{Tw}
\end{eqnarray}
while  the tachyon potential is given by
\begin{eqnarray}
V(T) &=& - \Lambda_5\ \mbox{sech}\left(\sqrt{-\frac{2}{3}\kappa_5^2\,\Lambda_5}\ T\right)
\sqrt{6\ \mbox{sech}^2\left(\sqrt{-\frac{2}{3}\kappa_5^2\,\Lambda_5}\ T\right)-1}\nonumber \\
&=& - \Lambda_5\ \sqrt{1+\mbox{sech}\left[\,H\,(2w+c)\right]} \sqrt{1+\frac{3}{2}\
\mbox{sech}\left[\,H\,(2w+c)\right]}, 
\label{VT}
\end{eqnarray}
where $H$, $c$ and $s>0$ are integration constants. It is worth mentioning that in the last two equations we have set
\begin{equation}
s=-\frac{6H^2}{\kappa_5^2\,\Lambda_5}
\label{s}
\end{equation}
with a negative bulk cosmological constant $\Lambda_5<0$.

Thus, the full solution corresponds to a non--trivial 5D configuration that describes a thick braneworld supported by a real tachyonic scalar field in which the induced metric on the 3--brane is described by a de Sitter 4D cosmological background.

It is evident that the warp factor decays and asymptotically vanishes along the extra dimension, whereas the tachyon scalar possesses a kink or antikink--like profile. In contrast with previously found solutions for this system, see for instance \cite{palkar}, the self--interaction tachyon potential $V(T)$ can be explicitly expressed in terms of the
tachyonic scalar field $T$ and has a maximum/minimum at the position of the brane, but
it is positive/negative definite as it can be seen from (\ref{VT}). Moreover, since the tachyon
field is bounded, the self--interaction potential $V(T)$ is bounded as well and remains always real.

The 5D curvature scalar for the solution (\ref{scalewarpfactors}) is given by
\begin{equation}
R=-\frac{14}{3}\kappa_5^2\,\Lambda_5\,\mbox{sech}\left[\,H\,(2w+c)\right].
\label{R5}
\end{equation}
It is worth noticing that this invariant is positive definite and vanishes asymptotically, yielding an asymptotically 5D Minkowski spacetime. This is a very peculiar and novel property of the braneworld configuration under consideration since, usually, in the absence of matter a negative cosmological constant $\Lambda_5$ leads to an asymptotically $AdS_5$ spacetime. However, from (\ref{accion}) it is clear that both the cosmological constant $\Lambda_5$ and the self--interaction potential $V(T)$ contribute to the asymptotically vanishing overall effective cosmological constant of the 5D spacetime of the tachyonic braneworld.


\section{Localization of gauge vector fields}

Now we shall investigate the localization and mass spectrum of gauge fields on the tachyonic thick braneworld previously reviewed.  Let us begin with the five--dimensional action of a gauge vector field
\begin{eqnarray}\label{action_Vector}
S_{1} = -\frac{1}{4}\int d^{5}x \sqrt{-g}~ g^{M N}
 g^{RS}F_{MR}F_{NS},
\end{eqnarray}
where $F_{MN}=\partial_{M}A_{N}-\partial_{N}A_{M}$. The corresponding field equations that follow from this action are
\begin{eqnarray}
\frac{1}{\sqrt{-g}} \partial_{M}\left(\sqrt{-g} g^{M N} g^{R S}
F_{NS}\right) = 0.
\end{eqnarray}
By recalling the background metric (\ref{metric_w}), the field equations can be expressed in terms of the hatted 4D metric $\hat{g}^{\mu\nu}$ as follows
\begin{eqnarray}
\label{fieldeqns}
 \frac{1}{\sqrt{-\hat{g}}}\partial_\nu\left(\sqrt{-\hat{g}} ~
      \hat{g}^{\nu \rho}\hat{g}^{\mu\lambda}F_{\rho\lambda}\right)
    +{\hat{g}^{\mu\lambda}}e^{-f}\partial_w
      \left(e^{f} F_{5\lambda}\right)  = 0, ~~\\
 \partial_\mu\left(\sqrt{-\hat{g}}~ \hat{g}^{\mu \nu} F_{\nu 5}\right) =
 0.~~
 \label{fieldeqn2}
\end{eqnarray}
Let us choose the gauge condition $A_5=0$ by using our freedom. Therefore, the action (\ref{action_Vector}) adopts the form
\begin{eqnarray}
S_1 = - \frac{1}{4} \int d^5 x \sqrt{-g} \bigg\{
        g^{\mu\alpha} g^{\nu\beta} F_{\mu\nu}F_{\alpha\beta}
        +2e^{-2f} g^{\mu\nu} \partial_w A_{\mu} \partial_w A_{\nu}
       \bigg\}.
\label{actionVector2}
\end{eqnarray}
By further making use of the following KK decomposition for the gauge field $A_{\mu}$
\begin{eqnarray}
A_{\mu}(x^{\lambda},w)=\sum_n
a^{(n)}_\mu(x^{\lambda})\rho_n(w)e^{-f/2}
\label{KKdecompositionOfAM}
\end{eqnarray} 
the action (\ref{actionVector2}) is expressed as
\begin{eqnarray}
S_1 = - \frac{1}{4} \sum_{m}\sum_{n}\int d^4 x\ dw \sqrt{-\hat{g}}\,
       \bigg[\hat{g}^{\mu\alpha} \hat{g}^{\nu\beta}f^{(m)}_{\mu\nu}f^{(n)}_{\alpha\beta}\;\rho_m\rho_n
       +2\,\hat{g}^{\mu\nu}a^{(m)}_{\mu}a^{(n)}_{\nu}\bigg(\rho'_m\rho'_n-\frac{1}{2}f'(\rho_m\rho_n)'+\frac{1}{4}f'^2\rho_m\rho_n
       \bigg)\bigg],
\label{actionVector3}
\end{eqnarray}
where $f^{(n)}_{\mu\nu} = \partial_\mu a^{(n)}_\nu - \partial_\nu a^{(n)}_\mu$ is the 4D field strength tensor and $'$ denotes derivatives with respect to the fifth dimension. Here we shall make use of the following relations 
$\rho'_m\rho'_n=(\rho'_m\rho_n)'-\rho''_m\rho_n$ and $f'(\rho_m\rho_n)'=(f'\rho_m\rho_n)'-f''\rho_m\rho_n$
in order to get rid of total derivative terms along the fifth dimension and get the action
\begin{eqnarray}
S_1 = - \frac{1}{4} \sum_{m}\sum_{n}\int d^4 x\ dw \sqrt{-\hat{g}}\,
       \bigg\{\hat{g}^{\mu\alpha} \hat{g}^{\nu\beta}f^{(m)}_{\mu\nu}f^{(n)}_{\alpha\beta}\;\rho_m\rho_n
       +2\,\hat{g}^{\mu\nu}a^{(m)}_{\mu}a^{(n)}_{\nu}\bigg[-\rho''_m+\left(\frac{1}{2}f''+\frac{1}{4}f'^2\right)\rho_m\bigg]\rho_n\bigg\}.
\label{actionVector4}
\end{eqnarray}
By further applying the orthonormalization condition for the extra dimensional functions $\rho_m(w)$
\begin{eqnarray}
 \int^{\infty}_{-\infty}  \;\rho_m(w)\rho_n(w)dw=\delta_{mn},
 \label{normalizationCondition2}
\end{eqnarray}
the action (\ref{actionVector4}) transforms into
\begin{eqnarray}
S_1 = \sum_{n}\int d^4 x \sqrt{-\hat{g}}~
       \bigg( - \frac{1}{4}\hat{g}^{\mu\alpha} \hat{g}^{\nu\beta}
             f^{(n)}_{\mu\nu}f^{(n)}_{\alpha\beta}
       - \frac{1}{2}m_{n}^2 ~\hat{g}^{\mu\nu}
           a^{(n)}_{\mu}a^{(n)}_{\nu}
       \bigg),
\label{actionVector5}
\end{eqnarray}
where, since the perfactor of $\hat{g}^{\mu\nu}a^{(m)}_{\mu}a^{(n)}_{\nu}$ must define a squared mass for the gauge potential $a^{(n)}_{\mu}$, we have assumed that the KK vector modes $\rho_n(w)$ satisfy the following Schr\"{o}dinger equation
\begin{eqnarray}
   \left[-\partial^2_w +V_1(w) \right]{\rho}_n(w)=m_n^{2}   {\rho}_n(w)
    \label{SchEqVector1}
\end{eqnarray}
with an analog quantum mechanical potential $V_{1}(w)$ given by
\begin{eqnarray}
\label{V_Vector}
 V_{1}(w)= \frac{1}{2}\partial^2_w f + \frac{1}{4}\left(\partial_w f\right)^2 = \frac{H^{2}}{4}\left[1-5\, \text{sech}^{2}(2Hw)\right],
\end{eqnarray}
where we have used the warp factor from (\ref{scalewarpfactors}) (with $c=0$) in the second equality. 

Alternatively, by making use of the KK decomposition (\ref{KKdecompositionOfAM}), the Schr\"{o}dinger equation (\ref{SchEqVector1}) can be obtained from the field equation (\ref{fieldeqns}) upon the corresponding definition of the 4D squared mass $m_n^{2}$ for the vector field, in agreement with the previous approach. Thus, from (\ref{fieldeqns}) under the KK decomposition (\ref{KKdecompositionOfAM}) we get
\begin{eqnarray}
 \frac{1}{\sqrt{-\hat{g}}}\partial_\nu \bigg(
\sqrt{-\hat{g}}\,\hat{g}^{\mu\alpha} \hat{g}^{\nu\beta}\sum_{n}f^{(n)}_{\alpha\beta}\bigg)\rho_n
       +\hat{g}^{\mu\lambda}\sum_{n}a^{(n)}_{\lambda}\bigg[\rho''_n-\left(\frac{1}{2}f''+\frac{1}{4}f'^2\right)\rho_n\bigg]=0.
\label{fieldeqnrho}
\end{eqnarray}
We note that the first term defines the 4D mass of a massive gauge vector field according to
\begin{eqnarray}
 \frac{1}{\sqrt{-\hat{g}}}\partial_\nu \bigg(
\sqrt{-\hat{g}}\,\hat{g}^{\mu\alpha} \hat{g}^{\nu\beta}f^{(n)}_{\alpha\beta}\bigg)=m_n^2\hat{g}^{\mu\lambda}a^{(n)}_{\lambda},
\label{fieldeqnrho}
\end{eqnarray}
giving rise to the Schr\"{o}dinger equation (\ref{SchEqVector1}) for the extra dimensional profile $\rho_n(w)$. It should be stated as well that the remaining field equation (\ref{fieldeqn2}) leads to the Lorentz gauge condition for the 4D gauge vector field $a^{(n)}_{\mu}$.
   
In order to study the localization of the gauge vector field we observe that the quantum mechanical potential (\ref{V_Vector}) has a minimum equal to $-H^{2}$ at $w=0$ and asymptotically adopts a positive definite value ${H^{2}}/{4}$ at $w=\pm\infty$, guaranteeing the existence of a mass gap in the corresponding spectrum. Thus, equation (\ref{SchEqVector1}) with the quantum mechanical potential (\ref{V_Vector}) is nothing else than the Schr\"{o}dinger equation with a modified P\"{o}schl-Teller potential
\begin{eqnarray}\label{SchEqVector2}
\left[-\partial_{u}^{2}-\frac{5}{16}\text{sech}^{2}u\right]\rho_{n}=E_{n}\rho_{n},
\end{eqnarray}
where we have set $u=2Hw$ and $E_{n}=\frac{m_{n}^{2}}{4H^{2}}-\frac{1}{16}$. The corresponding to (\ref{SchEqVector2}) energy spectrum is found to be $E_{n}=-\left(\frac{1}{4}-n\right)^{2}$, where $n$ is an integer satisfying $0\leq n <\frac{1}{4}$. Therefore, there is just one bound state in the energy spectrum of the KK vector profile, namely, the massless one  $\rho_{0}$ with energy $E_{0}=-1/16$. Thus, the normalized zero mode corresponding to the ground state reads
\begin{eqnarray}
\label{ZeroVector}
\rho_{0}(w)=\frac{\sqrt{H}\left(\pi/2\right)^{1/4}}{2\,\Gamma\left(5/4\right)}\text{sech}^\frac{1}{4}(2Hw).
\end{eqnarray}

The continuous spectrum of massive vector KK modes starts at $m^{2}={H^{2}}/{4}$ and asymptotically turns into plane waves according to the traditional quantum mechanical picture. Since these modes are not localized on the brane, when the energy of vector modes is larger than $H/2$, they leak into the extra dimension.

Let us see how this comes about. The general solution for the Schr\"odinger equation (\ref{SchEqVector1}) can be expressed as a linear combination of associated Legendre functions of first and second kind of degree $1/4$ and order $\mu=i\,\sigma=\sqrt{\frac{1}{16}-\frac{m^2}{4H^2}}$:
\be
\rho_n(w)=k_1\ P_{1/4}^{\mu} \left[\tanh(2Hw)\right]+k_2\ Q_{1/4}^{\mu}\left[\tanh(2Hw)\right],
\label{Psi_m} 
\ee 
where $k_1$ and $k_2$ are integration constants. As pointed out above, the continuous spectrum of gauge eigenfunctions starts at $2m>H$, when the order becomes imaginary, and asymptotically describe plane waves according to \cite{PRD0709.3552}:
\begin{eqnarray}
\rho^{\mu}_{\pm}(w)= C_{\pm}(\sigma)P^{\pm i\sigma}_{1/4} \Bigl(\tanh\bigl(2Hw\bigr)\Bigr)
\sim \frac{C_{\pm}(\sigma)}{\Gamma(1\mp i\sigma)}e^{\pm i2H\sigma w}
= \frac{1}{\sqrt{2\pi}}e^{\pm i2H\sigma w}. 
\end{eqnarray}

\section{Conclusions and discussion}
\label{SecConclusion}

We have shown that gauge field localization in the de Sitter thick braneworld generated by gravity coupled to a tachyonic scalar field is feasible since the normalizable vector zero mode can be localized on the brane. The spectrum consists of a massless zero mode (the ground state) and a series of continuous massive KK gauge modes separated by a mass gap from the massless one, as it also happens in the KK spectrum of the metric fluctuations \cite{germanetal}. The mass gap is proportional to $H$, and hence, proportional to the square root of the 4D cosmological constant. Thus, the height of the mass gap, which determines the energy scale at which the KK fluctuations can be excited, somehow is defined by the 4D cosmological constant and is big/small if the Hubble parameter is big/small. 

It would be interesting to consider the localization of fermion fields in the tachyonic braneworld model under consideration. This study could lead to the possibility of computing the corrections to Coulomb's law due to the KK massive modes of the gauge vector field within the model. It is also interesting to perform the analysis of one loop corrections to brane-to-brane propagators as in \cite{kirpichnikov} in order to study possible divergent pathologies related to the existence of the mass gap of the model under consideration. These two issues are currently under consideration, and we do hope to overcome the difficulties we found and report on them in the near future.

\section*{Acknowledgement}

\noindent The authors acknowledge useful and illuminating discussions with Ra\'ul Henr\'\i quez. AHA is grateful to A.S. Adamidou for her valuable support. AHA and ADR are grateful to MCTP, UNACH for hospitality, and thank SNI for support. ADR acknowledges a postdoctoral grant from CONACYT while the research of  AHA was supported from grant \textquotedblleft Programa de Apoyo a Proyectos de Investigaci\'on e Innovaci\'on Tecnol\'ogica\textquotedblright\, (PAPIIT) UNAM, IN103413-3, {\it Teor\'ias de Kaluza-Klein, 
inflaci\'on y perturbaciones gravitacionales.}

\end{document}